\documentclass[]{svjour3}  
\usepackage{epsfig} 
\usepackage{graphicx}
\usepackage{dcolumn}
\usepackage{bm}
\usepackage[usenames]{color}
\usepackage{ulem} 
\usepackage{amsmath}
\usepackage{amssymb}
\usepackage[]{natbib}

\journalname{Bulletin of Mathematical Biology}

\begin{document}

\title{Predicting unobserved exposures from seasonal epidemic data}

\author{Eric Forgoston \and Ira B. Schwartz} 

\institute{Eric Forgoston \at
      Department of Mathematical Sciences, 
      Montclair State University,
      1 Normal Avenue,
      Montclair, NJ 07043, USA\\
      \email{eric.forgoston@montclair.edu}
\and Ira B. Schwartz \at
      Nonlinear Systems Dynamics Section,
      Plasma Physics Division,
      Code 6792,
      U.S. Naval Research Laboratory,
      Washington, DC 20375, USA
      \email{ira.schwartz@nrl.navy.mil}
}

\date{Received: date / Accepted: date}

\maketitle

\begin{abstract}
We consider a stochastic Susceptible-Exposed-Infected-Recovered (SEIR)
epidemiological model with a contact rate that fluctuates seasonally.  Through
the use of a nonlinear, stochastic projection,
we are able to analytically determine the lower dimensional manifold on which
the deterministic and stochastic dynamics correctly interact.  Our method
produces a low dimensional stochastic model that captures the same timing of
disease outbreak and the same amplitude and phase of recurrent behavior seen in the
high dimensional model.  Given seasonal epidemic data consisting of the number
of infectious individuals, our method enables a data-based model
prediction of the number of unobserved exposed individuals over very long times.
\end{abstract}

\section{Introduction}
As a dynamical process in epidemics, noise increasingly plays an important role
when using models to understand and  predict disease outbreaks. Stochastic
effects figure  prominently in finite populations, which can range from 
ecological dynamics~\citep{Marion2000} to childhood epidemics in
cities~\citep{Rohani2002,Nguyen2008}.  For populations 
with seasonal forcing, noise  comes into play in the prediction of large
outbreaks~\citep{Rand1991,BillingsBS02,stolhu07}.  External random
perturbations, {such as those arising from random migrations~\citep{amp2007},} change the probabilistic
prediction of epidemic outbreak amplitudes as well as their
control~\citep{Schwartz2004}. 

Epidemic stochasticity
may arise from several sources, such as internal interactions due to random
contacts in finite well-mixed
populations~\citep{Nasell1999,Doering2005}, as well as contacts on
{structured contact}
networks~\citep{shasch08}. On the other hand, it may arise from
random external fluctuations, such as immigration perturbations~\citep{Anderson91,amp2007}. In many cases, whether noise arises  externally or
internally, it  may have
far 
ranging effects on outbreak predictability. For example, noise impacts predictability  in climate
and disease~\citep{ISI:000256349900004},  produces random-looking outbreaks from its interaction with mass
action terms~\citep{NR2002}, and  
amplifies the peaks of epidemics due to its interaction with  resonant
frequencies~\citep{amp2007}. Other areas of noise research 
have been postulated as the source of chaos in epidemics~\citep{Rand1991,bisc02},
whereby noise explicitly interacts with the  underlying  topology of the
epidemics model. As a direct  consequence of its topological interaction,
random chaotic-like switching may occur between
small amplitude and large outbreaks when including the effects of stochasticity~\citep{BillingsBS02}. 

To connect with data, time series analysis from spatio-temporal cases has been done to
generate parameters  for use in epidemic models.  Tools from nonlinear time series analysis of 
measles data~\citep{schaffer93,Blarer1999} have pointed to the important
realization that any mathematical  models might need to capture chaotic
behavior, which is deterministic and predictable only in the short term. As a result, time series data analysis has
centered on the assumption that a deterministic processes dominates
the time series, but quantifying determinism from the statistics may
be inconclusive. In~\cite{Tid93}, analysis specifically
points out that complex, or chaotic behavior may be detected, even
when it is not deterministic. For example, sensitive dependence on initial
conditions in short, noisy, non-chaotic time series, such as epidemic data, may be indicated
by positive Lyapunov exponents. 

Another time series analysis method which also includes noise is that of Time-series
Susceptible-Infected-Recovered (TSIR) modeling~\citep{BFG2002}. Here the authors do
local fits of time-series data from England and Wales to get measures of local
reproductive rates of infection. The main assumption is  that in the
pre-vaccine years, all newborns introduced into the population as susceptible
individuals become infected. However, as constructed the model lacks
predictive power since the parameter fits, which include noise parameters,
are local in time. It also excludes any latency period of infection, since it
only considers models of SIR type. 

Connections of data with full models which are
higher dimensional are difficult if one includes other relevant
epidemiological parameters and realistic noise. {Coupled patch models of
  cities have limited data time series when compared to the large simulation model dimension. Such limited data sets imply the need for accurate lower
dimensional models to reduce parameter unknowns.} Latency of infection, which
introduces a series of  exposed classes {approximating mean delay times}, is
one example which {generates high dimensional models}, but is
{omitted} in modeling diseases used in the TSIR approach. However, it is known
rigorously that the  dynamics in higher dimensional deterministic models
relaxes asymptotically onto lower dimensional hyper-surfaces~\citep{ss83,shbisc07}. The dynamics in these reductions are purely
deterministic, and rely on nonlinear center manifold reduction methods. The advantage
in  doing center manifold reductions is that if one can only observe certain components
of a disease, then it is possible to explicitly construct a function which
relates the unobserved components (such as latency, or {asymptomatic} infections), to
those explicitly measured, or observed. The current state-of-the-art, however, now shows 
stochastic model reduction must be done correctly in order to connect with
observed data~\citep{rob08}.  When examining models based on observed data, it will improve prediction by  examining full
stochastic models which include all epidemiological factors, and then reduce
them properly to lower dimensions, whereby noise is projected properly. 

The purpose of this paper is to examine
a method of nonlinear, stochastic projection so that the deterministic
and stochastic dynamics interact correctly on the lower-dimensional
manifold and predict correctly the  {outbreak} dynamics when compared to the full
system when fitted to data.  In a previous paper~\citep{fobisc09},  we showed {how} noise affects the
timing of outbreaks {for a time independent system}. Therefore, it was concluded that  it is essential
to produce a low-dimensional system which captures the correct timing
of the outbreaks as well as the amplitude and phase of any recurrent
behavior {for any given measured realization of an outbreak}. 

For stochastic model reduction, there exist several potential methods
for general problems.  For systems with certain spectral requirements,
the existence of a stochastic center manifold was proven in~\cite{box89}.
Non-rigorous stochastic normal form analysis (which leads to the stochastic
center manifold) was performed in~\cite{knowie83,coelti85,nam90,namlin91}, as
well as others~\citep{arn98,arnimk98}.  In~\cite{rob08},
the construction of the stochastic
normal form coordinate transform is transparent, so we used this method
to derive the reduced stochastic center manifold equation when there is no
seasonal forcing using standard parameters~\citep{fobisc09}.

Here we take a different approach, and consider the data to fit a predictive
model of SEIR type, and then perform stochastic reduction in a time dependent
system. Rather than perform local temporal fits of a model to data, the data
is fit over a {long} period of time {compared to the infectious
period}. Then, given the
parameters and noise, we {explicitly construct} the stochastic manifold. The dynamics which
lives on the manifold allows us to predict the unobserved exposed, yet not
infectious, individuals explicitly. 


\section{SEIR Model}
We begin by describing a stochastic version of the SEIR model found
in~\cite{ss83}.  It is assumed that a given population can be divided into
four classes, each of which evolves in time.  The classes are defined as follows:
\begin{enumerate}
\item The class of susceptible individuals is denoted by $s(t)$.  Each susceptible individual may contract the disease from an infectious individual.
\item The class of exposed individuals is denoted by $e(t)$.  Each exposed
  individual has been infected by the disease but is not yet infectious.
\item The class of infectious individuals is denoted by $i(t)$.  Each infectious
  individual is capable of transmitting the disease to a susceptible
  individual.
\item The class of recovered individuals is denoted by $r(t)$.  Each recovered
  individual is immune to the disease.
\end{enumerate}
We also assume that the population size, denoted by $N$, is constant so that
$s(t)+e(t)+i(t)+r(t)=N$.  Denoting $S(t)=s(t)/N$, $E(t)=e(t)/N$,
$I(t)=i(t)/N$, and $R(t)=r(t)/N$, then the population class variables $S(t)$,
$E(t)$, $I(t)$, and $R(t)$ represent fractions of the total population and
$S(t)+E(t)+I(t)+R(t)=1$.  In terms of these new variables, the governing
equations of our stochastic SEIR model are given as
\begin{subequations}
\begin{flalign} 
\dot{S}(t)&=\mu -\beta (t) I(t)S(t) -\mu S(t),\label{e:Sdot}\\
\dot{E}(t)&=\beta (t) I(t)S(t) -(\alpha +\mu)E(t),\label{e:Edot}\\
\dot{I}(t)&=\alpha E(t) -(\gamma +\mu)I(t)+\sigma I(t)\phi(t),\label{e:Idot}\\
\dot{R}(t)&=\gamma I(t) -\mu R(t),\label{e:Rdot}
\end{flalign}
\end{subequations}
where $\sigma$ is the standard deviation of the noise intensity
$D=\sigma^2/2$, and $\phi$ is a stochastic white noise term that is
characterized by the following correlation functions:
\begin{equation} 
\langle\phi(t)\rangle=0,\label{e:mean}
\end{equation}
\begin{equation} 
\langle\phi(t)\phi(t^{\prime})\rangle=\delta(t-t^{\prime}).
\end{equation}
We only consider multiplicative noise in the infectives since that is the one
quantity that is measurable {as a function of cases.  However, since not every person who contracts a
disease will be treated by a doctor, and since some doctors may neglect to
consistently file reports with the monitoring agencies, the data is inherently
noisy.  In addition, 
restricting the noise to observations renders the problem easier to understand
analytically.} \footnote[1]{We note the fact that while the inclusion of noise terms on
other components makes the analysis more difficult, it will not affect the
predictions as long as we stay away from bifurcation points.} {Equations~(\ref{e:Sdot})-(\ref{e:Rdot}) and all of the other stochastic
differential equations that follow are interpreted in the Stratonovich sense.}

In Eqs.~(\ref{e:Sdot})-(\ref{e:Rdot}), the birth and death rate are described
by $\mu$, the rate of infection is described by $\alpha$, and the rate of
recovery is described by $\gamma$.  Additionally, the contact rate
$\beta (t)$ fluctuates seasonally, and we have chosen to represent $\beta$
with the following two-harmonic sinusoidal forcing function:
\begin{equation}
\beta (t) = \beta_0 [1 + \beta_1\cos{(2\pi t + \zeta_1)} + \beta_2\cos{(2\pi
  t\omega +\zeta_2)}].\label{e:beta}
\end{equation}

It should be noted that if one neglects the $\sigma I(t)\phi(t)$ stochastic
term, then the deterministic form of Eqs.~(\ref{e:Sdot})-(\ref{e:Rdot}) is
such that $S+E+I+R=1$.  In the stochastic problem, the four components will
not necessarily sum to unity {due to fluctuations in the total population}.  But since the noise has zero mean, on average
the total population will remain close to unity.  This fact, along with the
fact that $R$ is decoupled from Eqs.~(\ref{e:Sdot})-(\ref{e:Idot}), allows us to assume that the
dynamics are described sufficiently by Eqs.~(\ref{e:Sdot})-(\ref{e:Idot}).

\section{Deterministic Model Reduction}
We will reduce the dimension of the system given by
Eqs.~(\ref{e:Sdot})-(\ref{e:Idot}) using deterministic center manifold
theory.  The analysis begins by neglecting the $\sigma I(t)\phi(t)$ term and
considering the {autonomous} SEIR model {in the absence of periodic drive
  so that $\beta_1=\beta_2=0$}.  There are two {steady states} of the
deterministic system.  The first {steady state} corresponds to a disease-free, or
extinct, equilibrium state and is given as
\begin{equation}
\label{e:fixed_pt_mort_df}
(S_e,E_e,I_e)=(1,0,0).
\end{equation}
The second {steady state} corresponds to an endemic equilibrium state and is given as
\begin{equation}
\label{e:fixed_pt_mort}
(S_0,E_0,I_0)=\left (
{\frac {  1 }{R_0 }},
{\frac {(\gamma +\mu)   }{\alpha }}I_0 ,
{\frac {\mu }{ \beta_0 }}\left (R_0 -1\right )
  \right ), 
\end{equation}
where
\begin{equation}
\label{e:R0}
R_0=\frac{\alpha\beta_0}{\left (\gamma +\mu\right )\left (\alpha +\mu\right )}.
\end{equation}
Biologically, $R_0$ is interpreted as the basic reproductive rate and gives the
number of secondary cases produced by a lone infectious individual in a
population of susceptible individuals during one infectious period. From here
on, we assume that $R_0 > 1$ so that an endemic equilibrium always exists.

A general nonlinear system may be transformed so that the system's
linear part has a block diagonal form consisting of three matrix blocks.  The
first matrix block will possess eigenvalues with positive real part; the
second matrix block will possess eigenvalues with negative real part; and the
third matrix block will possess eigenvalues with zero real part.  These three
matrix blocks are respectively associated with the unstable eigenspace, the
stable eigenspace, and the center eigenspace.  If there are no eigenvalues
with positive real part, then the orbits will rapidly decay to the center
eigenspace.

Equations~(\ref{e:Sdot})-(\ref{e:Idot}) can not be written in a block
diagonal form with one matrix block possessing eigenvalues with negative real
part and the other matrix block possessing eigenvalues with zero real part.
{Even though it is possible to construct a center manifold from a system
  not in separated block form~\citep{chilat97}, it is much easier to apply the center
  manifold theory to a system with separated stable and center directions.}
Therefore, we transform the original system given by
Eqs.~(\ref{e:Sdot})-(\ref{e:Idot}) to a new system of equations that will have
the eigenvalue structure that is needed to apply center manifold theory.  The
theory allows one to find an invariant center manifold that passes through a
fixed point and to which one can restrict the new transformed system. 
\subsection{Transformation of the SEIR Model}
To ease the analysis, we define a new set of variables, $\bar{S}$, $\bar{E}$, and $\bar{I}$, as
$\bar{S}(t)=S(t)-S_0$, $\bar{E}(t)=E(t)-E_0$, and $\bar{I}(t)=I(t)-I_0$.
These new variables are substituted into Eqs.~(\ref{e:Sdot})-(\ref{e:Idot}). 

Then, treating $\mu$ as a small parameter, we rescale time by letting $t=\mu\tau$.
We may then introduce the following rescaled parameters:  $\alpha=\alpha_0/\mu$ and $\gamma=\gamma_0/\mu$,
where $\alpha_0$ and $\gamma_0$ are $\mathcal{O}(1)$.  The inclusion of the
parameter $\mu$ as a new state variable means that the terms in our rescaled
system which contain $\mu$ are now nonlinear terms.  Furthermore, the system
is augmented with the auxiliary equation $\frac{d\mu}{d\tau}=0$.  The addition
of this auxiliary equation contributes an extra simple zero eigenvalue to the
system and adds one new center direction that has trivial dynamics.  The shifted
and rescaled, augmented system of equations is given as follows: 
\begin{subequations}
\begin{flalign}
\frac{d\bar{S}}{d\tau}=& -\beta(t)\mu \bar{I}\bar{S} - \frac{\beta(t)(\alpha_0 +\mu^2)(\gamma_0
 +\mu^2)}{\alpha_0\beta_0}\bar{I} - \left [\frac{\alpha_0\mu^3 \beta(t)}{(\alpha_0 +\mu^2)(\gamma_0
    +\mu^2)} + \mu^2 -\frac{\mu^2\beta(t)}{\beta_0}\right ]\bar{S}
+\nonumber\\
&\left [\mu^2-\frac{\mu^2\beta(t)}{\beta_0} -\frac{\mu(\alpha_0 +\mu^2)(\gamma_0
 +\mu^2)}{\alpha_0\beta_0} + \frac{\mu\beta(t)(\alpha_0 +\mu^2)(\gamma_0
 +\mu^2)}{\alpha_0\beta_0^2}\right],\label{e:Sbardot_mort}\\
\frac{d\bar{E}}{d\tau}=&\beta(t)\mu \bar{I}\bar{S} + \frac{\beta(t)(\alpha_0 +\mu^2)(\gamma_0
 +\mu^2)}{\alpha_0\beta_0}\bar{I} +\frac{\mu^2\beta(t)[\alpha_0\beta_0\mu -(\alpha_0 +\mu^2)(\gamma_0
 +\mu^2)]}{\beta_0(\alpha_0 +\mu^2)(\gamma_0 +\mu^2)}\bar{S} - \nonumber\\
&(\alpha_0
+\mu^2) \bar{E}
+\frac{\mu(\alpha_0+\mu^2)(\gamma_0+\mu^2)(R_0-1)}{\alpha_0\beta_0^2}(\beta(t)-\beta_0),\label{e:Ebardot_mort}
\end{flalign}
\begin{flalign}
\frac{d\bar{I}}{d\tau}=&\alpha_0 \bar{E} - (\gamma_0 +\mu^2) \bar{I},\label{e:Ibardot_mort}\\
\frac{d\mu}{d\tau}=&0\label{e:mudot},
\end{flalign}
\end{subequations}
where the endemic fixed point is now located at the origin.

The transformed system given by Eqs.~(\ref{e:Sbardot_mort})-(\ref{e:mudot}) is a
non-autonomous system due to the $\beta(t)$ term.  We generate the
corresponding autonomous system by replacing the cosine terms in
Eq.~(\ref{e:beta}) as follows: 
\begin{subequations}
\begin{flalign}
x_1=&\bar{x}_1=\cos{(2\pi t+\zeta_1)},\label{e:x1}\\
x_2=&\bar{x}_2=\cos{(2\pi\omega t+\zeta_2)}.\label{e:x2}
\end{flalign}
\end{subequations}
The autonomous system consists of Eqs.~(\ref{e:Sbardot_mort})-(\ref{e:mudot})
plus the following four additional equations:
\begin{subequations}
\begin{flalign}
\frac{d\bar{x}_1}{d\tau} =& \mu\bar{x}_1 -2\pi\mu\bar{x}_3 -\mu\bar{x}_1\left
  (\bar{x}_3^2+\bar{x}_1^2\right ),\label{e:dx1}\\
\frac{d\bar{x}_2}{d\tau} =& \mu\bar{x}_2 -2\pi\omega\mu\bar{x}_4 -\mu\bar{x}_2\left
  (\bar{x}_2^2+\bar{x}_4^2\right ),\label{e:dx2}\\
\frac{d\bar{x}_3}{d\tau} =& \mu\bar{x}_3 +2\pi\mu\bar{x}_1 -\mu\bar{x}_3\left
  (\bar{x}_3^2+\bar{x}_1^2\right ),\label{e:dx3}\\
\frac{d\bar{x}_4}{d\tau} =& \mu\bar{x}_4 +2\pi\omega\mu\bar{x}_2 -\mu\bar{x}_4\left
  (\bar{x}_2^2+\bar{x}_4^2\right ),\label{e:dx4}
\end{flalign}
\end{subequations}
where the specific form of the right-hand side of
Eqs.~(\ref{e:dx1})-(\ref{e:dx4}) corresponds to a limit cycle. 

The Jacobian of Eqs.~(\ref{e:Sbardot_mort})-(\ref{e:mudot}) and Eqs.~(\ref{e:dx1})-(\ref{e:dx4}) is computed to
zeroth-order in $\mu$ and is evaluated
at the origin.  Ignoring the $\mu$ and $\bar{x}_i$ components, the Jacobian has only two linearly independent eigenvectors. Therefore, the
Jacobian is not diagonalizable.  However, it is possible to
transform Eqs.~(\ref{e:Sbardot_mort})-(\ref{e:Ibardot_mort}) to a block diagonal
form with {a separated} eigenvalue structure.  {As mentioned
  previously, this block structure makes the center manifold
analysis easier.}  We use a transformation matrix, ${\bf P}$, consisting of the
two linearly independent eigenvectors of the Jacobian along with a third
vector chosen to be linearly independent.  There are many choices for this
third vector; our choice is predicated on keeping the vector as simple as
possible.  This transformation matrix is given as follows:
\begin{equation}
\setlength{\arraycolsep}{4pt}
  \renewcommand{\arraystretch}{0.8}
{\bf P} = \left [ \begin{array}{cccccccc}
1 & & & 1 & & & &0 \\
& & & & & & &\\
-\frac{\alpha_0+\gamma_0}{\gamma_0} & & & 0 & & & &0 \\
& & & & & & &\\
\frac{\alpha_0+\gamma_0}{\gamma_0} & & & 0 & & & &1  \\
 \end{array} \right ].
\end{equation}

Using the fact that
$(\bar{S},\bar{E},\bar{I})^T = {\bf P}\cdot (U,V,W)^T$, then the transformation matrix leads to the following definition of new variables, $U$, $V$, and $W$:
\begin{subequations}
\begin{flalign}
U =& \frac{-\gamma_0}{\alpha_0 + \gamma_0}\bar{E},\label{e:U}\\
V =& \bar{S} + \frac{\gamma_0}{\alpha_0 + \gamma_0}\bar{E},\label{e:V}\\
W =& \bar{I} + \bar{E}.\label{e:W}
\end{flalign}
\end{subequations}
The application of the transformation matrix to
Eqs.~(\ref{e:Sbardot_mort})-(\ref{e:Ibardot_mort}) leads to the 
transformed evolution equations
\begin{subequations}
\begin{flalign}
\frac{dU}{d\tau} =& F_1(U,V,W,\mu),\label{e:dU}\\
\frac{dV}{d\tau} =& F_2(U,V,W,\mu),\label{e:dV}\\
\frac{dW}{d\tau} =& F_3(U,V,W,\mu),\label{e:dW}\\
\frac{d\mu}{d\tau} =&0, \label{e:dmu}
\end{flalign}
\end{subequations}
where $F_1$, $F_2$, and $F_3$ are complicated expressions given in Appendix~\ref{sec:determ-model-red}.

\subsection{Center Manifold Analysis}
As we did previously, Eqs.~(\ref{e:dU})-(\ref{e:dmu}) may be written in
autonomous form by replacing the cosine terms in $\beta(t)$ with
Eqs.~(\ref{e:x1})-(\ref{e:x2}) and expanding the system to include Eqs.~(\ref{e:dx1})-(\ref{e:dx4}).
The Jacobian of Eqs.~(\ref{e:dU})-(\ref{e:dmu}) and
Eqs.~(\ref{e:dx1})-(\ref{e:dx4}) to zeroth-order in
$\mu$ and evaluated at the origin is
\begin{equation}
\setlength{\arraycolsep}{4pt}
  \renewcommand{\arraystretch}{0.8}
\left [ \begin{array}{cc|ccccccccccccccccccccc}
-(\alpha_0+\gamma_0) & & & &0 & & & -\frac{\gamma_0^2}{(\alpha_0+\gamma_0)}&
& & 0 &&& 0 &&& 0 &&& 0 &&& 0\\
& & & & & & & & & &\\
\hline
& & & & & & & & & &\\
0 & & & &0 & & & -\frac{\alpha_0\gamma_0}{(\alpha_0+\gamma_0)}& & & 0 &&& 0 &&& 0 &&& 0 &&& 0\\
& & & & & & & & & &\\
0 & & & &0 & & & 0 & & & 0 &&& 0 &&& 0 &&& 0 &&& 0\\
& & & & & & & & & &\\
0 & & & &0 & & & 0 & & & 0 &&& 0 &&& 0 &&& 0 &&& 0\\
& & & & & & & & & &\\
0 & & & &0 & & & 0 & & & 0 &&& 0 &&& 0 &&& 0 &&& 0\\
& & & & & & & & & &\\
0 & & & &0 & & & 0 & & & 0 &&& 0 &&& 0 &&& 0 &&& 0\\
& & & & & & & & & &\\
0 & & & &0 & & & 0 & & & 0 &&& 0 &&& 0 &&& 0 &&& 0\\
& & & & & & & & & &\\
0 & & & &0 & & & 0 & & & 0 &&& 0 &&& 0 &&& 0 &&& 0\end{array} \right ],
\end{equation} 
which shows that Eqs.~(\ref{e:dU})-(\ref{e:dmu}) and
Eqs.~(\ref{e:dx1})-(\ref{e:dx4}) may be rewritten in the form
\begin{flalign}
&\frac{d{\bf x}}{d\tau}= {\bf A}{\bf x} + {\bf f}({\bf x},{\bf  y},\mu),\label{e:CMformx}\\
&\frac{d{\bf y}}{d\tau}= {\bf B}{\bf y} + {\bf g}({\bf x},{\bf  y},\mu),\label{e:CMformy}\\
&\frac{d\mu}{d\tau}=\frac{dx_i}{d\tau}=0,
\end{flalign}
where ${\bf x}=(U)$, ${\bf y}=(V,W)$, ${\bf A}$ is a constant matrix with
eigenvalues that have negative real parts, ${\bf B}$ is a constant matrix with
eigenvalues that have zero real parts, and ${\bf f}$ and ${\bf g}$ are
nonlinear functions in ${\bf x}$, ${\bf y}$ and $\mu$.  In particular,
\begin{equation}
{\bf A}=\left [ \begin{array}{c}
-(\alpha_0 +\gamma_0)\end{array} \right ], \,\,\,\,\,\,\,\,\,\,\,\,\,\,\, {\bf B}=\left [
\begin{array}{cccc}
0 &&& -\frac{\alpha_0\gamma_0}{(\alpha_0 +\gamma_0)}\\
&&&\\
0 &&& 0\end{array} \right ].
\end{equation}

Therefore, this new system of equations which is an exact transformation of
Eqs.~(\ref{e:Sdot})-(\ref{e:Idot}) will rapidly collapse onto a lower-dimensional manifold
 given by center manifold theory~\citep{car81,chilat97,dulusc03}. Furthermore, since ${\bf x}$ is
 associated with ${\bf A}$ and ${\bf y}$ is associated with ${\bf B}$, we know that the center manifold is given by
\begin{equation}
\label{e:CM}
U=h(V,W,\mu,x_i),
\end{equation}
where $h$ is an unknown function. 

Substitution of Eq.~(\ref{e:CM}) into Eq.~(\ref{e:dU}) leads to the following
center manifold condition:
\begin{small}
\begin{flalign}
\label{e:CMcond}
&\frac{\partial h}{\partial V}\frac{dV}{d\tau}+\frac{\partial
  h}{\partial W}\frac{dW}{d\tau}={\frac {\mu\,\gamma_0\, \left( \beta(t)-\beta_0 \right) 
 \left[ -\beta_0\,\mu\,\alpha_0+ \left( \alpha_0+{\mu}^{2}
 \right)  \left( \gamma_0+{\mu}^{2} \right)  \right] }{{\beta_0
}^{2}\alpha_0\, \left( \alpha_0+\gamma_0 \right) }}+ &&\nonumber\\
&\left( -\alpha_0-{\mu}^{2}-{\frac { \left( \alpha_0+{\mu}^{2}
 \right)  \left( \gamma_0+{\mu}^{2} \right)
 \beta(t)}{\beta_0\,\alpha_0}}+{\frac {{\mu}^{2}\gamma_0\,\beta(t)}{\beta_0\,
 \left( \alpha_0+\gamma_0 \right) }}- {\frac {\beta(t)\,{\mu}^{3}\gamma_0\,\alpha_0}{ \left( \alpha_0+{\mu
}^{2} \right)  \left( \gamma_0+{\mu}^{2} \right)  \left( \alpha_0+\gamma_0
\right) }} \right) h+\nonumber\\
& \left( -{\frac {\beta(t)\,{\mu}^{3}\gamma_0\,\alpha_0}{ \left( \alpha_0+
{\mu}^{2} \right)  \left( \gamma_0+{\mu}^{2} \right)  \left( \alpha_0+\gamma_0 \right) }}+{\frac {{\mu}^{2}\gamma_0\,\beta(t)}{\beta_0\, \left( \alpha_0+\gamma_0 \right) }}
 \right) V-{\frac {\gamma_0\,\beta(t)\, \left( \gamma_0+{\mu}^{2} \right)  \left(
      \alpha_0+{\mu}^{2} \right) W}{\alpha_0\, \left( \alpha_0+\gamma_0
    \right) \beta_0}}-\nonumber\\
&{\frac
  {\gamma_0\,\beta(t)\,\mu\,W\,h}{\alpha_0+\gamma_0}}-\beta(t)\,\mu\,h\,V-{\frac {\gamma_0\,\beta(t)\,\mu\,V\,W}{\alpha_0+\gamma_0}}-\beta(t)\,\mu\,{h}^{2}.
\end{flalign}
\end{small}
In general, it is not possible to solve the center manifold condition for the
unknown function, $h(V,W,\mu,x_i)$.  Therefore, a Taylor series expansion of
$h(V,W,\mu,x_i)$ in $V$, $W$, $\mu$, and $x_i$ is
substituted into the center manifold equation.  The unknown coefficients are
determined by equating terms of the same order, and the center manifold
equation is found to be
\begin{equation}
\label{e:CMeq}
U=-\frac{\gamma_0^2}{(\alpha_0+\gamma_0)^2}W + \mathcal{O}(\epsilon^3),
\end{equation}
where $\epsilon =|(V,W,\mu)|$ so that $\epsilon$ provides a count of the
number of $V$, $W$, and $\mu$ factors in any one term.

{Substitution of the center manifold equation [Eq.~(\ref{e:CMeq})] into
  Eqs.~(\ref{e:dV}) and~(\ref{e:dW}) leads to the reduced system of evolution
  equations that describes the dynamics on the center manifold.  One can solve
this reduced system of equation for $V$ and $W$, and then use
Eq.~(\ref{e:CMeq}) to find $U$.  In order to find the original $S$, $E$, and
$I$ variables, one can use the following relations between
the transformed variables $U$,
$V$, and $W$ and the original
$S$, $E$, and $I$ variables:}
\begin{subequations}
\begin{flalign}
S=&U+V+\frac{\left (\gamma +\mu\right )\left (\alpha +\mu\right )}{\alpha\beta_0},\label{e:S}\\
E=&-\frac{\left (\alpha +\gamma\right )}{\gamma}U + \frac{\left (\gamma
    +\mu\right )\mu}{\alpha\beta_0}\left (R_0 -1\right ),\label{e:E}\\
I=&\frac{\left (\alpha +\gamma\right )}{\gamma}U + W +\frac{\mu}{\beta_0}\left (R_0 -1\right ).\label{e:I}
\end{flalign}
\end{subequations}

\section{Stochastic Model Reduction}
Having found the deterministic center manifold equation, we now return to the
stochastic SEIR system given by Eqs.~(\ref{e:Sdot})-(\ref{e:Idot}).  We
transform the stochastic SEIR system using the same procedure as for the
deterministic system described in the previous section.  {As a result, we find the noise to first order in $\mu$ is independent of
the parametric drive $\beta (t)$, as evidenced in  Eqs.~(\ref{e:dUstoch})-(\ref{e:dWstoch}).} The only difference
is the effect of the transformation on the stochastic term.  The original
stochastic system contains one multiplicative noise term in one equation
[Eq.~(\ref{e:Idot})].  The new transformed stochastic system contains a linear
combination of two multiplicative and one additive noise terms.

In general, if there are stochastic terms associated with each of the
equations which comprise the original system, then the transformed system will
contain multiple additive and multiplicative noise terms.  In this case, all
of the additive noise terms in each equation can be considered as a new
additive noise term with a variance different from the original noise process.
In this situation, previous work~\citep{fobisc09} has shown that one should use a
normal form coordinate transform reduction method to properly project the
noise and dynamics onto the lower-dimensional manifold.  
Reference~\citep{fobisc09} outlines a general theory
that compares two methods to perform a stochastic model reduction: (i)
  deterministic center manifold method and (ii) stochastic normal form
  coordinate transform method.   When the
stochastic normal form coordinate transform reveals noise terms at low
order, then the
deterministic center manifold reduction cannot be applied, since the deterministic reduction
ignores the important noise terms, resulting in imperfect stochastic projection. On the other hand, if the stochastic normal
form coordinate transform yields noise at sufficiently high order, stochastic contributions are
negligible, and therefore a deterministic reduction may be used to perform the
  projection onto the low dimensional manifold. It should be pointed out that even when one uses the
  deterministic center manifold result, the result remains a stochastic one.

For this model, since the transformation yields only one additive noise term
which cannot be combined with the multiplicative terms to consider a new noise
process with a different variance, we can use the deterministic center
manifold result to reduce the stochastic model.  Additionally, we have
  explicitly computed the normal form coordinate transformation.  The result
  shows that the stochastic terms occur at high order, thus justifying our use
of the deterministic center manifold for this particular model.  Substituting the deterministic center manifold
equation given by Eq.~(\ref{e:CMeq}) into the full system of stochastic,
transformed equations gives the 
reduced stochastic model that
describes the dynamics on the center manifold. Since they are complicated, the specific forms of
the complete, stochastic transformed system and its associated reduced,
stochastic system are provided in Appendix~\ref{sec:stoch-model-red}.

\section{Results}
We numerically integrate the complete, stochastic system of transformed
equations of the SEIR model along with the reduced system of equations of the
SEIR model using a stochastic fourth-order Runge-Kutta integrator with a
constant time step size.  The complete system is solved for $U$, $V$, and $W$, while the reduced system
is solved for $V$, and $W$.  In this latter case, $U$ is estimated using the
center manifold equation given by Eq.~(\ref{e:CMeq}).   After the values of
$U$, $V$, and $W$ are known, the values of $S$, $E$, and $I$ are computed
using the transformations given by Eqs.~(\ref{e:S})-(\ref{e:I}).  

\begin{figure}
\begin{center}
\includegraphics[width=12cm]{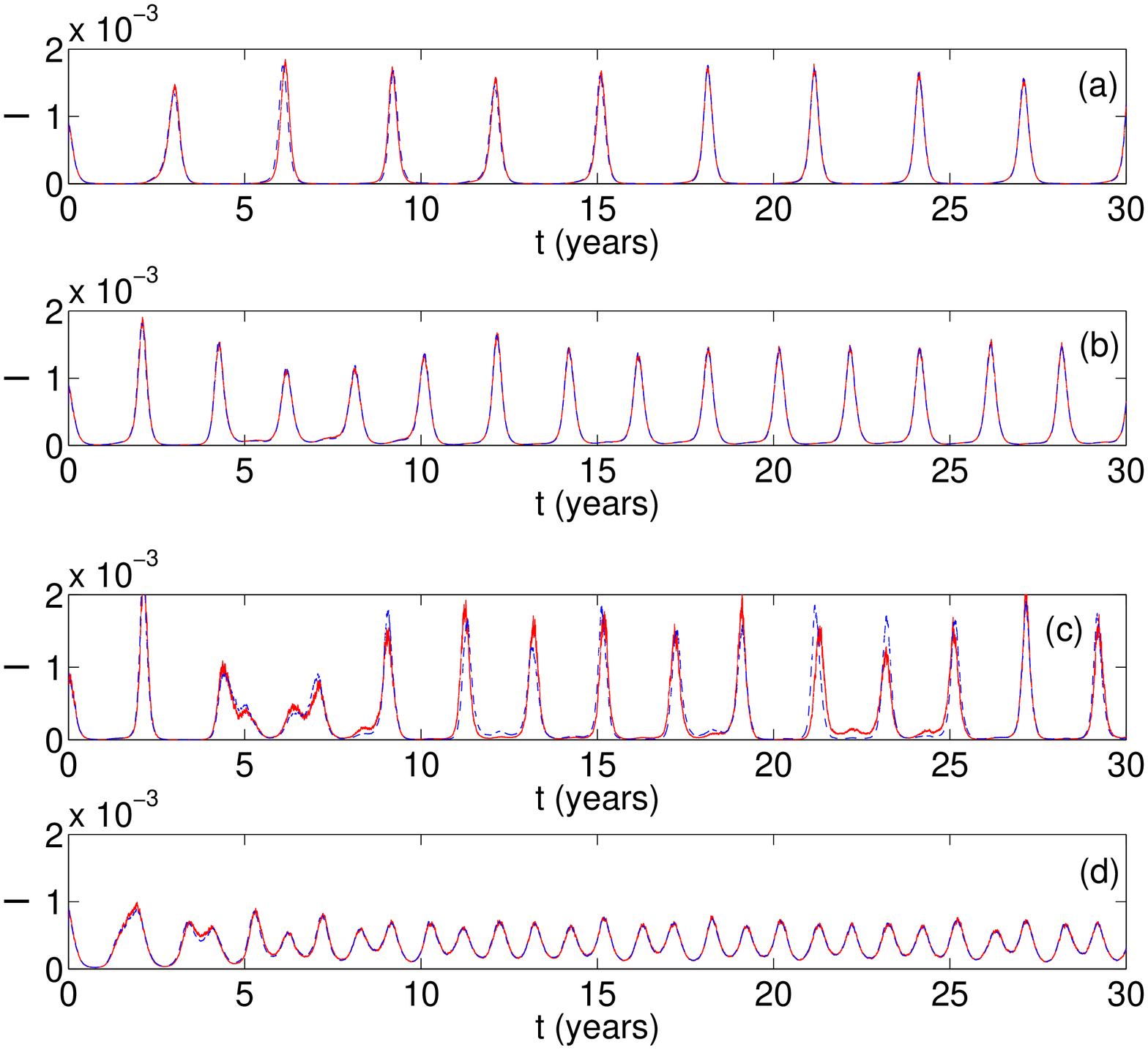}
\caption{\label{fig:TS_trans_full_and_reducedCM_tdbeta} {\bf Time
  series of the fraction of the population that is infected with a disease,
  $I$.}  The time series are found using the complete, stochastic system of
  transformed equations of the SEIR model (red, solid line) as well as the
  reduced, stochastic system of equations (blue, dashed line).  The parameter values used in the
  simulation are given as (a) $\mu =0.02({\rm  year})^{-1}$, $\alpha =1/0.0279({\rm  year})^{-1}$, $\gamma =1/0.01({\rm  year})^{-1}$, $\beta_0=1575.0({\rm  year})^{-1}$, $\beta_1 = 0.1$, $\beta_2=\zeta_1=\zeta_2=\omega=0$, and
$\sigma =2.0$; (b) the same as in (a) except now $\mu =0.03({\rm  year})^{-1}$; (c) the same as
in (a) except now $\mu =0.03({\rm  year})^{-1}$, $\beta_1 = 0.15$, and
$\sigma =5.0$; (d) the same as in (a) except now $\mu =0.04({\rm  year})^{-1}$.}
\end{center}
\end{figure}

Figures~\ref{fig:TS_trans_full_and_reducedCM_tdbeta}(a)-(d) compares two
time series of the fraction of the population that is infected with a
disease, $I$.  The first time series was computed using the complete,
stochastic system of transformed equations, while the second
time series was computed using the reduced, stochastic system of equations.
In Fig.~\ref{fig:TS_trans_full_and_reducedCM_tdbeta}(a), the following parameter values are used in the
computation: $\mu =0.02 ({\rm  year})^{-1}$, $\alpha =1/0.0279 ({\rm
  year})^{-1}$, $\gamma =1/0.01 ({\rm year})^{-1}$, $\beta_0=1575.0 ({\rm
  year})^{-1}$, $\beta_1 = 0.1$, $\beta_2=\zeta_1=\zeta_2=\omega=0$, and
$\sigma =2.0$.   These disease parameters correspond to typical measles
values~\citep{ss83,bisc02}.  There is excellent agreement between the two solutions shown in
Fig.~\ref{fig:TS_trans_full_and_reducedCM_tdbeta}(a).  The initial disease
outbreak is correctly captured by the reduced model.  Furthermore, the reduced
model correctly predicts outbreaks for a time scale on the order of decades.

Additionally, the solution found using the reduced, stochastic system agrees
very well with the solution found using the complete, stochastic system for a
wide range of parameter values.  The agreement can be seen in
Figs.~\ref{fig:TS_trans_full_and_reducedCM_tdbeta}(b)-(d).  By changing $\mu$,
$\beta_1$, and $\sigma$, one can obtain different frequency and amplitude
structure of the solutions.  Regardless, the reduced system still properly
captures the initial disease outbreak as well as the recurring outbreaks.  

\section{Comparison with Data}
\begin{figure}
\begin{center}
\includegraphics[width=12cm]{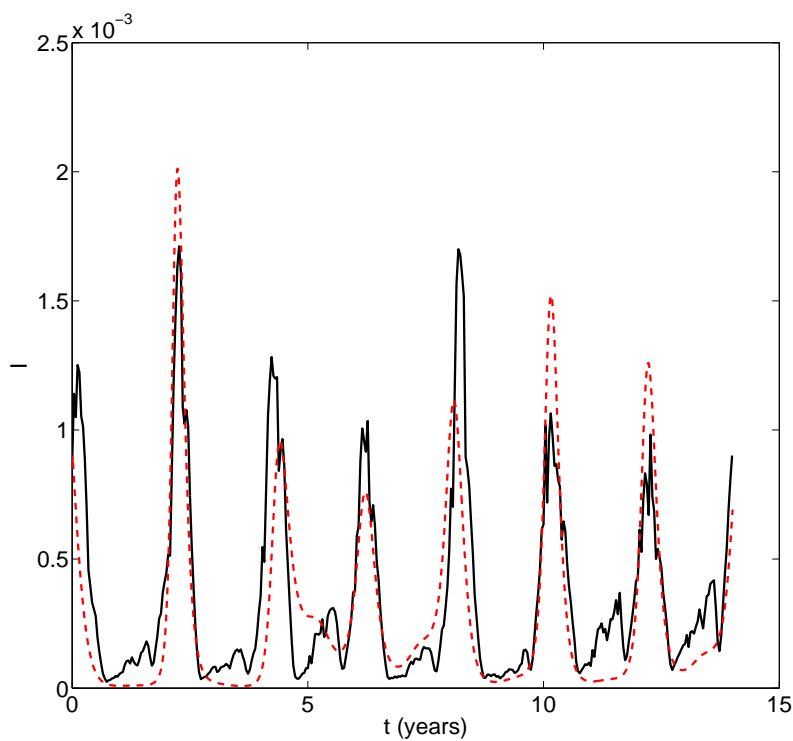}
\caption{\label{fig:data_rcm_determ} {\bf Time
  series of the fraction of the population that is infected with a disease,
  $I$.}  One time series is given by data~\citep{fine1982a} from 60 cities
  in England and Wales over the time span of 1953-1966 (black, solid line), while the
  other time series is computed using the reduced, stochastic system of
  equations (red, dashed line) for $\sigma=0.0$ (deterministic).  The cross-correlation between the two time
  series is
  $0.817$.  The parameter values used in the
  simulation are given as follows:   $\mu =0.0299({\rm  year})^{-1}$, $\alpha =30.0({\rm  year})^{-1}$,
  $\gamma =90.0({\rm  year})^{-1}$, $\beta_0=1329.345({\rm  year})^{-1}$, $\beta_1 = 0.1$, $\beta_2=0.05$,
  $\zeta_1=-0.2128$, $\zeta_2=-0.2554$, and $\omega=0.521$.}
\end{center}
\end{figure}

Measles have been registered on a weekly basis via mandatory notification in
the UK~\citep{fine1982a}, with a reporting rate in the pre-vaccination era
better than fifty percent~\citep{clarkson1985}.
We use existing measles data~\citep{fine1982b}  that consists of the number of infectious
individuals from 60 cities in England and Wales over the 14 year time span of 1953-1966.  The
data from all of these cities is aggregated so that the total population is 14,602,896.
Figure~\ref{fig:data_rcm_determ} shows the time series of the aggregate data as the
fraction of the population that is infected with the disease.  Fitting the
data with the deterministic version of Eqs.~(\ref{e:Sdot})-(\ref{e:Idot})
yields the following disease parameter values:  $\mu =0.0299 ({\rm
  year})^{-1}$, $\alpha =30.0 ({\rm  year})^{-1}$,
  $\gamma =90.0 ({\rm  year})^{-1}$, $\beta_0=1329.345 ({\rm  year})^{-1}$, $\beta_1 = 0.1$, $\beta_2=0.05$,
  $\zeta_1=-0.2128$, $\zeta_2=-0.2554$, and $\omega=0.521$.
  Using these parameter values, the reduced system of equations is solved with
  $\sigma =0.0$ (deterministic case), and the resulting time series is also
  shown in Fig.~\ref{fig:data_rcm_determ}.

One can see that the agreement between the two time series is quite good.  In
particular, the reduced system accurately captures the timing of each of the
major outbreaks.  We also have computed the cross-correlation of the two time
series shown in Fig.~\ref{fig:data_rcm_determ} to be approximately $0.817$.
The cross-correlation measures the similarity between the two time series.  If the time series were identical, the cross-correlation would be equal to
$1.0$.  Although the time series of Fig.~\ref{fig:data_rcm_determ} are not
identical, their high cross-correlation value quantitatively suggests good
agreement between the measured data and the computed time series.

\begin{figure}
\begin{center}
\includegraphics[width=12cm]{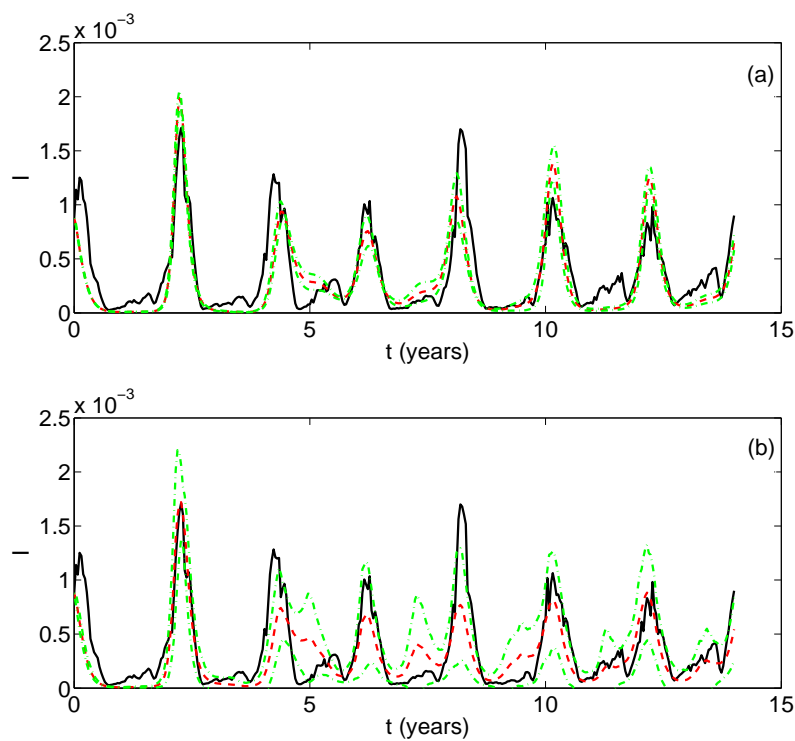}
\caption{\label{fig:data_rcm_stoch} {\bf Time
  series of the fraction of the population that is infected with a disease,
  $I$.}  One time series is given by data~\citep{fine1982a} (black, solid line), while the
  other time series is an average solution computed using the reduced, stochastic system of
  equations (red, dashed line) using $25$ realization of noise with intensity
  (a) $\sigma =2.0$, and (b) $\sigma =10.0$.  The range of $I$ within one
  standard deviation of the mean also is denoted (green, dashed-dotted line).
 The cross-correlation between the data time series and the average time
  series from the reduced model in (a) is
  $0.810$, while the cross-correlation between the two time series in (b) is $0.655$.  The parameter values used in the
  simulation are given as follows:   $\mu =0.0299({\rm  year})^{-1}$, $\alpha =30.0({\rm  year})^{-1}$,
  $\gamma =90.0({\rm  year})^{-1}$, $\beta_0=1329.345({\rm  year})^{-1}$, $\beta_1 = 0.1$, $\beta_2=0.05$,
  $\zeta_1=-0.2128$, $\zeta_2=-0.2554$, and $\omega=0.521$.}
\end{center}
\end{figure}

The solution that is computed using the reduced, stochastic system is very
robust to noise.  The standard deviation of the noise intensity $\sigma$ must be fairly
large to significantly affect the accuracy of the computed solution.
Figure~\ref{fig:data_rcm_stoch}(a) compares the data time series with the
average time
series computed using the reduced, stochastic model with $25$ realizations of
noise with intensity $\sigma=2.0$.
Figure~\ref{fig:data_rcm_stoch}(b) is similar, except that $\sigma=10.0$ was
used for the reduced model computation.  Also shown in
Figs.~\ref{fig:data_rcm_stoch}(a)-(b) is the range of infectives that fall
within one standard deviation of the average solution.

By comparing Fig.~\ref{fig:data_rcm_determ} with
Fig.~\ref{fig:data_rcm_stoch}(a), one can see that the noise has not had a
great effect on the mean solution found using the reduced model.  In fact, the cross-correlation between the two time
series of Fig.~\ref{fig:data_rcm_stoch}(a) is approximately $0.810$, which is
very near the deterministic cross-correlation value.  Beyond $\sigma =2.0$ the
noise has a significant degrading effect on the reduced model solution.
Figure~\ref{fig:data_rcm_stoch}(b) shows that the overall agreement between the
two time series is still relatively good.  However, by comparing
Fig.~\ref{fig:data_rcm_stoch}(a) with Fig.~\ref{fig:data_rcm_stoch}(b), one
can see that the larger noise has led to significant differences between the
two reduced model solutions.  The cross-correlation value is $0.655$, which is
much lower than the cross-correlation found for lower values of $\sigma$.

\begin{figure}
\begin{center}
\includegraphics[width=12cm]{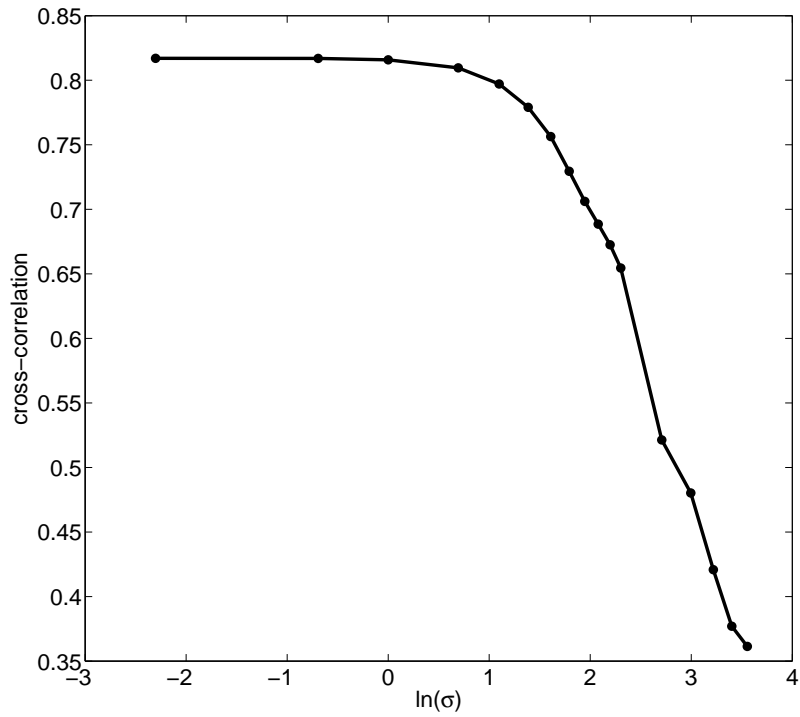}
\caption{\label{fig:cross-corr} {\bf Cross-correlation between the data time series
  and the time series which is computed using the reduced, stochastic system of
  equations for noise
intensities ranging from $\sigma=0.1$ to $\sigma=35.0$.}  The cross-correlation for $\sigma =0.0$ (deterministic) is
  $0.817$ (data point is not shown).  For each value of $\sigma$, the
  data point represents the average of individual cross-correlations computed
  using $25$ realizations of the noise. }
\end{center}
\end{figure}

To obtain a comprehensive idea of the effect of the noise, we have computed
the cross-correlation between the data time series and the time series
computed using the reduced, stochastic system for noise
intensities ranging from $\sigma=0.1$ to $\sigma=35.0$.
Figure~\ref{fig:cross-corr} shows the cross-correlation as a function of
$\ln{(\sigma)}$.  

The reduction method also allows one to predict the unobserved number of
exposed individuals based on the observed number of infected individuals.
Using the center manifold equation given by Eq.~(\ref{e:CMeq}) along with the
equations which relate the
original $S$, $E$, and $I$ variables to the transformed $U$, $V$, and $W$
variables, one can find the following relation between exposed and infected
individuals:
\begin{equation}
\label{e:Ecm}
E=\frac{\gamma}{\alpha}I +\frac{\mu^2}{\alpha\beta_0}\left (R_0 -1\right ).
\end{equation}
In Eq.~(\ref{e:Ecm}), the first term on the right hand side provides a measure of the
statistical steady state flow conditions for exposure and infection, since $\gamma^{-1}$
and $\alpha^{-1}$ are respectively the compartmental recovery and infection times.  The
second term on the right hand side is a correction, not predicted in the
classical theory.  An increase in $R_0$
leads to an increase in the speed of infection per infectious period.  This
increase induces an increase in the number of exposed individuals. 

\begin{figure}
\begin{center}
\includegraphics[width=12cm]{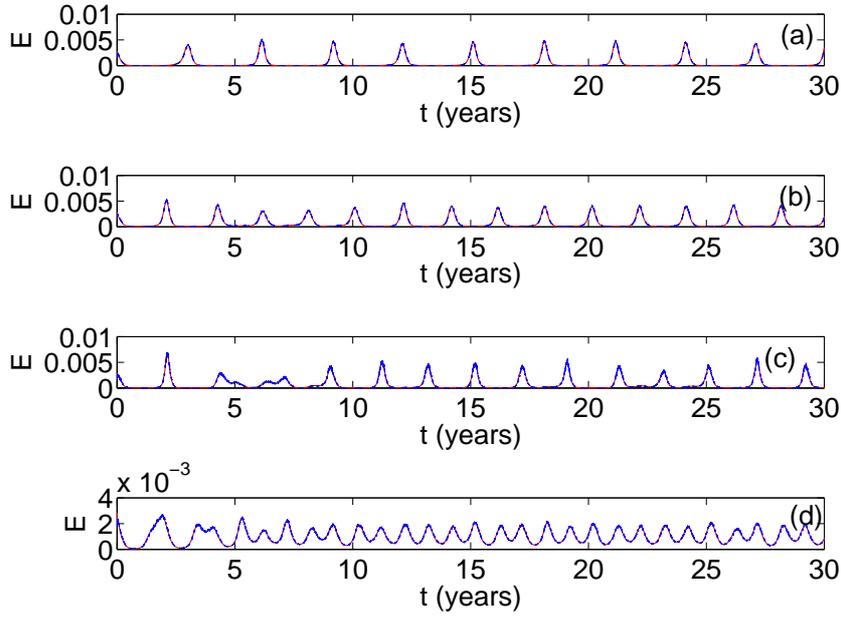}
\caption{\label{fig:Exposed_trans_full_and_CM_prediction} {\bf Time
  series of the fraction of the population that has been exposed to a disease,
  $E$.}  The time series are found using the complete, stochastic system of
  transformed equations of the SEIR model (red, solid line) as well as the
  predictive equation given by Eq.~(\ref{e:Ecm}) (blue, dashed line).  The parameter values used in the
  simulation are given as (a) $\mu =0.02({\rm  year})^{-1}$, $\alpha =1/0.0279({\rm  year})^{-1}$, $\gamma =1/0.01({\rm  year})^{-1}$, $\beta_0=1575.0({\rm  year})^{-1}$, $\beta_1 = 0.1$, $\beta_2=\zeta_1=\zeta_2=\omega=0$, and
$\sigma =2.0$; (b) $\mu =0.03({\rm  year})^{-1}$, $\alpha =1/0.0279({\rm  year})^{-1}$, $\gamma =1/0.01({\rm  year})^{-1}$, $\beta_0=1575.0({\rm  year})^{-1}$, $\beta_1 = 0.1$, $\beta_2=\zeta_1=\zeta_2=\omega=0$, and
$\sigma =2.0$; (c) $\mu =0.03({\rm  year})^{-1}$, $\alpha =1/0.0279({\rm  year})^{-1}$, $\gamma =1/0.01({\rm  year})^{-1}$, $\beta_0=1575.0({\rm  year})^{-1}$, $\beta_1 = 0.15$, $\beta_2=\zeta_1=\zeta_2=\omega=0$, and
$\sigma =5.0$; (d) $\mu =0.04({\rm  year})^{-1}$, $\alpha =1/0.0279({\rm  year})^{-1}$, $\gamma =1/0.01({\rm  year})^{-1}$, $\beta_0=1575.0({\rm  year})^{-1}$, $\beta_1 = 0.1$, $\beta_2=\zeta_1=\zeta_2=\omega=0$, and
$\sigma =2.0$.}
\end{center}
\end{figure}

Figures~\ref{fig:Exposed_trans_full_and_CM_prediction}(a)-(d) shows a
comparison between time series of the fraction of the population that has been
exposed to a disease, $E$.  The first time series was computed using the
complete, stochastic system of transformed equations.  We then used the
simulated $I$ data to predict the number of exposed individuals using
Eq.~(\ref{e:Ecm}).  The predicted number of exposed individuals constitutes
the second time series.  This comparison was performed for four different sets
of parameter values, corresponding to the values used in
Figs.~\ref{fig:TS_trans_full_and_reducedCM_tdbeta}(a)-(d).  As one can see,
there is excellent agreement between the predicted number of exposed
individuals and the actual, simulated number of exposed individuals.

\begin{figure}
\begin{center}
\includegraphics[width=12cm]{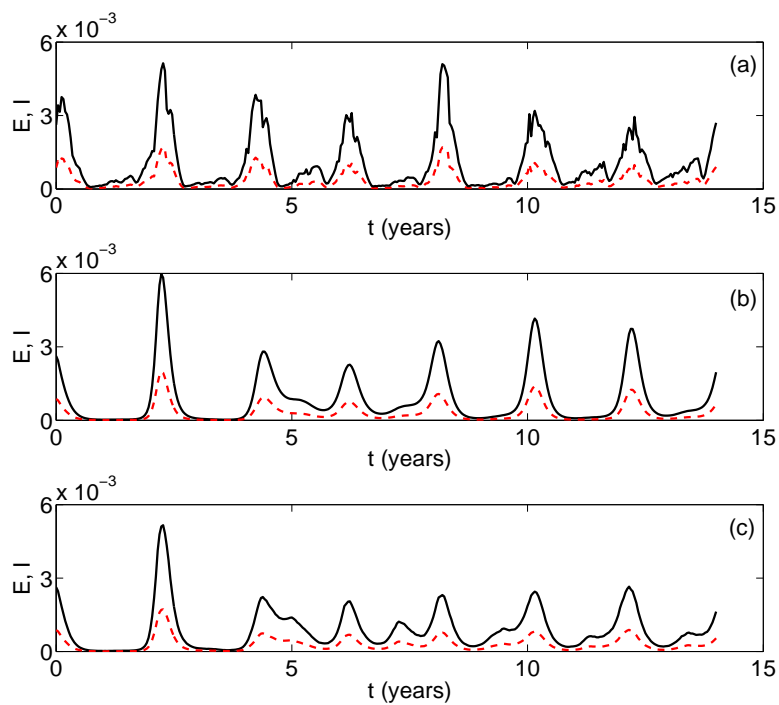}
\caption{\label{fig:exposed} {\bf Time
  series of the fraction of the population that has been exposed to a disease,
  $E$ (black, solid line) and infected with a disease,
  $I$ (red, dashed line).}  The time series of exposed individuals is based on
  the center manifold equation and is given by Eq.~(\ref{e:Ecm}).  Part (a)
  shows the observed data of infectious individuals along with the predicted,
  unobserved number of exposed individuals.  Parts (b) and (c) show the
  average number of infectious individuals computed using the reduced, stochastic system of
  equations using $25$ realizations of noise with intensity
  (b) $\sigma =2.0$, and (c) $\sigma =10.0$, along with the associated number
  of exposed individuals.  The parameter values used in the
  simulation are given as follows:   $\mu =0.0299({\rm  year})^{-1}$, $\alpha =30.0({\rm  year})^{-1}$,
  $\gamma =90.0({\rm  year})^{-1}$, $\beta_0=1329.345({\rm  year})^{-1}$, $\beta_1 = 0.1$, $\beta_2=0.05$,
  $\zeta_1=-0.2128$, $\zeta_2=-0.2554$, and $\omega=0.521$.}
\end{center}
\end{figure}

Figures~\ref{fig:exposed}(a)-(c) shows time series of the fraction of the
population that has been exposed to a disease, $E$, along with the fraction of
infectious individuals, $I$.  Fig.~\ref{fig:exposed}(a) shows the observed
data of infectious individuals and the predicted number of unobserved exposed
individuals.  Figs.~\ref{fig:exposed}(b)-(c) show the average number of
infectious individuals computed using the reduced, stochastic system of
equations using 25 realizations of noise for two different noise intensities
along with the associated number of exposed individuals.  

\section{Discussion}
We have considered a stochastic SEIR model where the contact rate fluctuates
seasonally and where multiplicative noise acts on the governing equation for
infectious individuals.  In this way, we emulate{d} the noise found within data of
measurable infectious individuals in a population.  The {main result} of our work
{was} the derivation of a lower-dimensional model whose solution, both in
amplitude and timing of outbreaks, agrees with the solution of the
higher-dimensional original model. 

There are many types of high-dimensional {stochastic} models {which lend themselves to
  model reduction}.  For example, a time delay is
often included in epidemic models when one wishes to model a disease
exposure time.  To reduce the analytical complications
introduced by the time delay, one can approximate the delay through a cascade
of hundreds of exposed compartments~\citep{mrv05}. Other high dimensional
models are generated when individual interactions within a population are
modeled as a network~\citep{Pastor-SatorrasV01,moreno2002}.  Researchers have considered epidemics on a variety of
static networks, including small world networks~\citep{Vazquez06} and transportation networks~\citep{ColizzaBBV06},
as well as on adaptive networks, where individuals may break their interaction
connections and ``rewire'' to form new interaction connections~\citep{shasch08}.

In these high dimensional model examples, one generally needs to resort to
massive computation and analytical results are usually very difficult or
impossible to obtain.  In particular, it is not currently possible to perform
these computations in real-time.  However, {many} high dimensional epidemic models
do contain time scales that are well separated.  Therefore, it is possible to
take advantage of these well-separated time scales to reduce the dimension of
the model.  In this article we have performed just such a model reduction.
The stochastic SEIR model with seasonal fluctuations{, which contains fast
  collapse and slow dynamic time scales,} illustrates the power of
our method. It is important to note that the analysis could
straightforwardly be extended to a SEIR-type model where the exposed class was
modeled using hundreds of compartments.

{The mathematical/computational techniques used here are general, and can be applied
to many population dynamics problems.}  Our analysis {started} by transforming the deterministic SEIR system of equations to a new
system of equations with a specific eigenvalue structure.  Employing center
manifold theory, we {were} able to find the reduced system of equations that
describes the dynamics on the lower dimensional manifold.  Due to the specific
nature of the noise, we can use the deterministic center manifold equation to reduce
the stochastic SEIR model.  The end result is a reduced stochastic model that
accurately captures the timing of the initial disease outbreak as well as the
timing of subsequent outbreaks for decades long {times
  compared to the infectious period}.  The solution
to the reduced stochastic model additionally agrees very well in amplitude
with the solution to the original high-dimensional model.  Moreover, the
reduced model is robust in that it accurately captures the timing and amplitude for a wide range of
parameter values.

{As a direct application to observations,} we have also used our deterministic model to fit actual measles data.  Once
the fitting parameters were determined we solved our reduced stochastic model
and compared the resulting solution with the data.  Beyond providing good
agreement with the data, we saw that a large noise intensity was needed before
the stochastic solution significantly deviated from the data.  In this way,
the stochastic solutions are very robust to noise.  We were able to further
identify the noise effects through cross-correlation computations.

Generally, the actual data that is measured in society is that of the number
of {observed cases}.  Not only are exposed individuals not measured,
an exposed individual often will not even know of the exposure and the
infection that is to come.  However, with our novel stochastic reduction
method, we are now able to predict how many unobserved exposed individuals
there are in a population based solely on the measurable number of infectious individuals.

In summary, a new method of stochastic model
reduction for an epidemiological model with seasonal fluctuations has been
performed.  By capturing both the timing of disease outbreak as well as the
amplitude of the outbreak for long temporal scales, our reduced model provides
impressive time series prediction.  By accurately modeling actual stochastic
disease data, we enable the application of novel control methods where the
timing of vaccine delivery and a disease outbreak is important.  Moreover, the
method is general, and may be extended to a variety of compartmental and
network models, including models with a high dimension.

\begin{acknowledgements}
The authors gratefully acknowledge support from the Office of Naval
   Research, and the National
   Institutes of Health.  E.F. is supported by Award Number N0017310-2-C007
   from the Naval Research Laboratory (NRL).  {I.B.S. was
supported by the NRL Base Research Program N0001412WX30002, and  by} Award
Number R01GM090204 from the National Institute Of General Medical Sciences.  The content is solely the responsibility of the authors and does not necessarily represent the official views of the National Institute Of General Medical Sciences or the
National Institutes of Health.
\end{acknowledgements}


 \appendix
\section{Deterministic Model Reduction}\label{sec:determ-model-red}
The $F_1$, $F_2$, and $F_3$ expressions found in
Eqs.~(\ref{e:dU})-(\ref{e:dW}) are given as follows: 

\begin{footnotesize}
\begin{subequations}
\begin{flalign}
F_1(U,V,W,\mu) =& {\frac {\mu\,\gamma_0\, \left( \beta(t)-\beta_0 \right) 
 \left[ -\beta_0\,\mu\,\alpha_0+ \left( \alpha_0+{\mu}^{2}
 \right)  \left( \gamma_0+{\mu}^{2} \right)  \right] }{{\beta_0
}^{2}\alpha_0\, \left( \alpha_0+\gamma_0 \right) }}+ &&\nonumber\\
&\left( -\alpha_0-{\mu}^{2}-{\frac { \left( \alpha_0+{\mu}^{2}
 \right)  \left( \gamma_0+{\mu}^{2} \right)
 \beta(t)}{\beta_0\,\alpha_0}}+{\frac {{\mu}^{2}\gamma_0\,\beta(t)}{\beta_0\,
 \left( \alpha_0+\gamma_0 \right) }}- {\frac {\beta(t)\,{\mu}^{3}\gamma_0\,\alpha_0}{ \left( \alpha_0+{\mu
}^{2} \right)  \left( \gamma_0+{\mu}^{2} \right)  \left( \alpha_0+\gamma_0
\right) }} \right) U+\nonumber\\
& \left( -{\frac {\beta(t)\,{\mu}^{3}\gamma_0\,\alpha_0}{ \left( \alpha_0+
{\mu}^{2} \right)  \left( \gamma_0+{\mu}^{2} \right)  \left( \alpha_0+\gamma_0 \right) }}+{\frac {{\mu}^{2}\gamma_0\,\beta(t)}{\beta_0\, \left( \alpha_0+\gamma_0 \right) }}
 \right) V-{\frac {\gamma_0\,\beta(t)\, \left( \gamma_0+{\mu}^{2} \right)  \left(
      \alpha_0+{\mu}^{2} \right) W}{\alpha_0\, \left( \alpha_0+\gamma_0
    \right) \beta_0}}-\nonumber\\
&{\frac
  {\gamma_0\,\beta(t)\,\mu\,W\,U}{\alpha_0+\gamma_0}}-\beta(t)\,\mu\,U\,V-{\frac {\gamma_0\,\beta(t)\,\mu\,V\,W}{\alpha_0+\gamma_0}}-\beta(t)\,\mu\,{U}^{2},\label{e:F1}
\end{flalign}
\begin{flalign}
F_2(U,V,W,\mu) =& {\frac {\mu\, \left( \beta(t)-\beta_0 \right)  \left[ -\beta_0\,\mu\,\alpha_0+ \left( \alpha_0+{\mu}^{2} \right) 
 \left( \gamma_0+{\mu}^{2} \right)  \right] }{{\beta_0}^{2}
 \left( \alpha_0+\gamma_0 \right) }}+ &&\nonumber\\
&\left( -{\frac { \left( 
\alpha_0+{\mu}^{2} \right)  \left( \gamma_0+{\mu}^{2} \right) 
\beta(t)}{\beta_0\,\gamma_0}}+{\frac {\alpha_0\,{\mu}^{
2}\beta(t)}{\beta_0\, \left( \alpha_0+\gamma_0 \right) 
}}+\alpha_0-{\frac {\beta(t)\,{\mu}^{3}{\alpha_0}^{2}}{
 \left( \alpha_0+{\mu}^{2} \right)  \left( \gamma_0+{\mu}^{2}
 \right)  \left( \alpha_0+\gamma_0 \right) }} \right) U
+ \nonumber\\
&\left( -{\mu}^{2}+{\frac {\alpha_0\,{\mu}^{2}\beta(t)}{\beta_0\, \left( \alpha_0+\gamma_0 \right) }}-{\frac {\beta(t)\,{\mu}^{3}{\alpha_0}^{2}}{ \left( \alpha_0+{\mu}^{2}
 \right)  \left( \gamma_0+{\mu}^{2} \right)  \left( \alpha_0+\gamma_0 \right) }} \right) V-{\frac { \left( \alpha_0+{
\mu}^{2} \right)  \left( \gamma_0+{\mu}^{2} \right) \beta(t)\,W}{\beta_0\,
\left( \alpha_0+\gamma_0 \right) }}-\nonumber\\
&{
\frac {\beta(t)\,\alpha_0\,\mu\,W\,U}{\alpha_0+\gamma_0}}-{\frac {\beta(t)\,\alpha_0\,\mu\,U\,V}{\gamma_0}}-{\frac {\beta(t)\,\alpha_0\,\mu\,V\,W}{\alpha_0+\gamma_0}}-{\frac {\beta(t)\,\alpha_0\,\mu\,{U}^{2}}{\gamma_0}},\label{e:F2}
\end{flalign}
\begin{flalign}
F_3(U,V,W,\mu)=& -{\frac {\mu\, \left( \beta(t)-\beta_0 \right)  \left[ -\beta_0\,\mu\,\alpha_0+ \left( \alpha_0+{\mu}^{2} \right) 
 \left( \gamma_0+{\mu}^{2} \right)  \right]
}{\alpha_0\,{\beta_0}^{2}}}+&&\nonumber\\
& \left( {\frac {\beta(t)\, \left( \alpha_0+{\mu}^
{2} \right)  \left( \gamma_0+{\mu}^{2} \right)  \left( \alpha_0+\gamma_0 \right) }{\beta_0\,\alpha_0\,\gamma_0}}-{
\frac {\beta(t)\,{\mu}^{2}}{\beta_0}}-\alpha_0-\gamma_0
+{\frac {{\mu}^{3}\beta(t)\,\alpha_0}{ \left( \alpha_0+{\mu
}^{2} \right)  \left( \gamma_0+{\mu}^{2} \right) }} \right) U+\nonumber\\
& \left( -{\frac {\beta(t)\,{\mu}^{2}}{\beta_0}}+{\frac {{
\mu}^{3}\beta(t)\,\alpha_0}{ \left( \alpha_0+{\mu}^{2}
 \right)  \left( \gamma_0+{\mu}^{2} \right) }} \right) V+{
\frac { \left( \gamma_0+{\mu}^{2} \right)  \left(
    \beta(t)\,\alpha_0+\beta(t)\,{\mu}^{2}-\beta_0\,\alpha_0 \right)
  W}{\beta_0\,\alpha_0}}+\nonumber\\
&\beta(t)\,\mu\,W\,U+{\frac {\beta(t)\,\mu\, \left( \alpha_0+\gamma_0
 \right) U\,V}{\gamma_0}}+\beta(t)\,\mu\,V
\,W+{\frac {\beta(t)\,\mu\, \left( \alpha_0+\gamma_0
 \right) {U}^{2}}{\gamma_0}}.&&\label{e:F3}
\end{flalign}
\end{subequations}
\end{footnotesize}

\newpage

\section{Stochastic Model Reduction}\label{sec:stoch-model-red}
The stochastic, transformed equations are given as follows:

\begin{footnotesize}
\begin{subequations}
\begin{flalign}
\frac{dU}{d\tau} =& {\frac {\mu\,\gamma_0\, \left( \beta(t)-\beta_0 \right) 
 \left[ -\beta_0\,\mu\,\alpha_0+ \left( \alpha_0+{\mu}^{2}
 \right)  \left( \gamma_0+{\mu}^{2} \right)  \right] }{{\beta_0
}^{2}\alpha_0\, \left( \alpha_0+\gamma_0 \right) }}+ &&\nonumber\\
&\left( -\alpha_0-{\mu}^{2}-{\frac { \left( \alpha_0+{\mu}^{2}
 \right)  \left( \gamma_0+{\mu}^{2} \right)
 \beta(t)}{\beta_0\,\alpha_0}}+{\frac {{\mu}^{2}\gamma_0\,\beta(t)}{\beta_0\,
 \left( \alpha_0+\gamma_0 \right) }}- {\frac {\beta(t)\,{\mu}^{3}\gamma_0\,\alpha_0}{ \left( \alpha_0+{\mu
}^{2} \right)  \left( \gamma_0+{\mu}^{2} \right)  \left( \alpha_0+\gamma_0
\right) }} \right) U+\nonumber\\
& \left( -{\frac {\beta(t)\,{\mu}^{3}\gamma_0\,\alpha_0}{ \left( \alpha_0+
{\mu}^{2} \right)  \left( \gamma_0+{\mu}^{2} \right)  \left( \alpha_0+\gamma_0 \right) }}+{\frac {{\mu}^{2}\gamma_0\,\beta(t)}{\beta_0\, \left( \alpha_0+\gamma_0 \right) }}
 \right) V-{\frac {\gamma_0\,\beta(t)\, \left( \gamma_0+{\mu}^{2} \right)  \left(
      \alpha_0+{\mu}^{2} \right) W}{\alpha_0\, \left( \alpha_0+\gamma_0
    \right) \beta_0}}-\nonumber\\
&{\frac
  {\gamma_0\,\beta(t)\,\mu\,W\,U}{\alpha_0+\gamma_0}}-\beta(t)\,\mu\,U\,V-{\frac {\gamma_0\,\beta(t)\,\mu\,V\,W}{\alpha_0+\gamma_0}}-\beta(t)\,\mu\,{U}^{2},\label{e:dUstoch}
\end{flalign}
\begin{flalign}
\frac{dV}{d\tau} =& {\frac {\mu\, \left( \beta(t)-\beta_0 \right)  \left[ -\beta_0\,\mu\,\alpha_0+ \left( \alpha_0+{\mu}^{2} \right) 
 \left( \gamma_0+{\mu}^{2} \right)  \right] }{{\beta_0}^{2}
 \left( \alpha_0+\gamma_0 \right) }}+ &&\nonumber\\
&\left( -{\frac { \left( 
\alpha_0+{\mu}^{2} \right)  \left( \gamma_0+{\mu}^{2} \right) 
\beta(t)}{\beta_0\,\gamma_0}}+{\frac {\alpha_0\,{\mu}^{
2}\beta(t)}{\beta_0\, \left( \alpha_0+\gamma_0 \right) 
}}+\alpha_0-{\frac {\beta(t)\,{\mu}^{3}{\alpha_0}^{2}}{
 \left( \alpha_0+{\mu}^{2} \right)  \left( \gamma_0+{\mu}^{2}
 \right)  \left( \alpha_0+\gamma_0 \right) }} \right) U
+ \nonumber\\
&\left( -{\mu}^{2}+{\frac {\alpha_0\,{\mu}^{2}\beta(t)}{\beta_0\, \left( \alpha_0+\gamma_0 \right) }}-{\frac {\beta(t)\,{\mu}^{3}{\alpha_0}^{2}}{ \left( \alpha_0+{\mu}^{2}
 \right)  \left( \gamma_0+{\mu}^{2} \right)  \left( \alpha_0+\gamma_0 \right) }} \right) V-{\frac { \left( \alpha_0+{
\mu}^{2} \right)  \left( \gamma_0+{\mu}^{2} \right) \beta(t)\,W}{\beta_0\,
\left( \alpha_0+\gamma_0 \right) }}-\nonumber\\
&{
\frac {\beta(t)\,\alpha_0\,\mu\,W\,U}{\alpha_0+\gamma_0}}-{\frac {\beta(t)\,\alpha_0\,\mu\,U\,V}{\gamma_0}}-{\frac {\beta(t)\,\alpha_0\,\mu\,V\,W}{\alpha_0+\gamma_0}}-{\frac {\beta(t)\,\alpha_0\,\mu\,{U}^{2}}{\gamma_0}},\label{e:dVstoch}
\end{flalign}
\begin{flalign}
\frac{dW}{d\tau} =& -{\frac {\mu\, \left( \beta(t)-\beta_0 \right)  \left[ -\beta_0\,\mu\,\alpha_0+ \left( \alpha_0+{\mu}^{2} \right) 
 \left( \gamma_0+{\mu}^{2} \right)  \right]
}{\alpha_0\,{\beta_0}^{2}}}+&&\nonumber\\
& \left( {\frac {\beta(t)\, \left( \alpha_0+{\mu}^
{2} \right)  \left( \gamma_0+{\mu}^{2} \right)  \left( \alpha_0+\gamma_0 \right) }{\beta_0\,\alpha_0\,\gamma_0}}-{
\frac {\beta(t)\,{\mu}^{2}}{\beta_0}}-\alpha_0-\gamma_0
+{\frac {{\mu}^{3}\beta(t)\,\alpha_0}{ \left( \alpha_0+{\mu
}^{2} \right)  \left( \gamma_0+{\mu}^{2} \right) }} \right) U+\nonumber\\
& \left( -{\frac {\beta(t)\,{\mu}^{2}}{\beta_0}}+{\frac {{
\mu}^{3}\beta(t)\,\alpha_0}{ \left( \alpha_0+{\mu}^{2}
 \right)  \left( \gamma_0+{\mu}^{2} \right) }} \right) V+{
\frac { \left( \gamma_0+{\mu}^{2} \right)  \left(
    \beta(t)\,\alpha_0+\beta(t)\,{\mu}^{2}-\beta_0\,\alpha_0 \right)
  W}{\beta_0\,\alpha_0}}+\nonumber\\
&\beta(t)\,\mu\,W\,U+{\frac {\beta(t)\,\mu\, \left( \alpha_0+\gamma_0
 \right) U\,V}{\gamma_0}}+\beta(t)\,\mu\,V
\,W+{\frac {\beta(t)\,\mu\, \left( \alpha_0+\gamma_0
 \right) {U}^{2}}{\gamma_0}}+\nonumber\\
&\mu\sigma\left ( \frac{(\alpha_0 +\gamma_0)}{\gamma_0}U+W+I_0\right )\phi,\label{e:dWstoch}
\end{flalign}
\end{subequations}
\end{footnotesize}

As discussed in the article, we can use the deterministic center
manifold result to reduce the stochastic model.  Substituting the deterministic center manifold
equation given by Eq.~(\ref{e:CMeq}) into the full system of stochastic,
transformed equations gives the following 
reduced stochastic model that describes the dynamics on the center manifold:

\begin{footnotesize}
\begin{subequations}
\begin{flalign}
\frac{dV}{d\tau}=&{\frac {\mu\, \left( {\beta(t)}-{\beta_0} \right)  \left[ -{\beta_0}\,\mu\,{\alpha_0}+ \left( {\alpha_0}+{\mu}^{2} \right) 
 \left( {\gamma_0}+{\mu}^{2} \right)  \right] }{{{\beta_0}}^{2}
 \left( {\alpha_0}+{\gamma_0} \right) }}+ \nonumber\\
&\left( -{\frac { \left( 
{\alpha_0}+{\mu}^{2} \right)  \left( {\gamma_0}+{\mu}^{2} \right) 
{\beta(t)}}{{\beta_0}\,{\gamma_0}}}+{\frac {{\alpha_0}\,{\mu}^{
2}{\beta(t)}}{{\beta_0}\, \left( {\alpha_0}+{\gamma_0} \right) 
}}+{\alpha_0}-{\frac {{\beta(t)}\,{\mu}^{3}{{\alpha_0}}^{2}}{
 \left( {\alpha_0}+{\mu}^{2} \right)  \left( {\gamma_0}+{\mu}^{2}
 \right)  \left( {\alpha_0}+{\gamma_0} \right) }} \right) {\left (-\frac{\gamma_0^2}{\left (\alpha_0 +\gamma_0\right )^2}W\right )}
+ \nonumber\\
&\left( -{\mu}^{2}+{\frac {{\alpha_0}\,{\mu}^{2}{\beta(t)}}{{\beta_0}\, \left( {\alpha_0}+{\gamma_0} \right) }}-{\frac {{\beta(t)}\,{\mu}^{3}{{\alpha_0}}^{2}}{ \left( {\alpha_0}+{\mu}^{2}
 \right)  \left( {\gamma_0}+{\mu}^{2} \right)  \left( {\alpha_0}+{
\gamma_0} \right) }} \right) {V}-{\frac { \left( {\alpha_0}+{
\mu}^{2} \right)  \left( {\gamma_0}+{\mu}^{2} \right) {\beta(t)}\,{
W}}{{\beta_0}\, \left( {\alpha_0}+{\gamma_0} \right) }}-\nonumber\\
&{
\frac {{\beta(t)}\,{\alpha_0}\,\mu\,{W}\,{\left (-\frac{\gamma_0^2}{\left (\alpha_0 +\gamma_0\right )^2}W\right )}}{{\alpha_0
}+{\gamma_0}}}-{\frac {{\beta(t)}\,{\alpha_0}\,\mu\,{\left (-\frac{\gamma_0^2}{\left (\alpha_0 +\gamma_0\right )^2}W\right )}\,{
V}}{{\gamma_0}}}-{\frac {{\beta(t)}\,{\alpha_0}\,\mu\,{\it 
We}\,{W}}{{\alpha_0}+{\gamma_0}}}-{\frac {{\beta(t)}\,{\alpha_0}\,\mu\,{{\left (-\frac{\gamma_0^2}{\left (\alpha_0 +\gamma_0\right )^2}W\right )}}^{2}}{{\gamma_0}}},
\end{flalign} 
\begin{flalign}
\frac{dW}{d\tau}=&-{\frac {\mu\, \left( {\beta(t)}-{\beta_0} \right)  \left( -{\beta_0}\,\mu\,{\alpha_0}+ \left( {\alpha_0}+{\mu}^{2} \right) 
 \left( {\gamma_0}+{\mu}^{2} \right)  \right)
}{{\alpha_0}\,{{\beta_0}}^{2}}}+\nonumber\\
& \left( {\frac {{\beta(t)}\, \left( {\alpha_0}+{\mu}^
{2} \right)  \left( {\gamma_0}+{\mu}^{2} \right)  \left( {\alpha_0
}+{\gamma_0} \right) }{{\beta_0}\,{\alpha_0}\,{\gamma_0}}}-{
\frac {{\beta(t)}\,{\mu}^{2}}{{\beta_0}}}-{\alpha_0}-{\gamma_0}
+{\frac {{\mu}^{3}{\beta(t)}\,{\alpha_0}}{ \left( {\alpha_0}+{\mu
}^{2} \right)  \left( {\gamma_0}+{\mu}^{2} \right) }} \right) {\it 
\left (-\frac{\gamma_0^2}{\left (\alpha_0 +\gamma_0\right )^2}W\right )}+\nonumber\\
& \left( -{\frac {{\beta(t)}\,{\mu}^{2}}{{\beta_0}}}+{\frac {{
\mu}^{3}{\beta(t)}\,{\alpha_0}}{ \left( {\alpha_0}+{\mu}^{2}
 \right)  \left( {\gamma_0}+{\mu}^{2} \right) }} \right) {V}+{
\frac { \left( {\gamma_0}+{\mu}^{2} \right)  \left( {\beta(t)}\,{
\alpha_0}+{\beta(t)}\,{\mu}^{2}-{\beta_0}\,{\alpha_0} \right) {
W}}{{\beta_0}\,{\alpha_0}}}+\nonumber\\
&{\beta(t)}\,\mu\,{W}\,{\it 
\left (-\frac{\gamma_0^2}{\left (\alpha_0 +\gamma_0\right )^2}W\right )}+{\frac {{\beta(t)}\,\mu\, \left( {\alpha_0}+{\gamma_0}
 \right) {\left (-\frac{\gamma_0^2}{\left (\alpha_0 +\gamma_0\right )^2}W\right )}\,{V}}{{\gamma_0}}}+{\beta(t)}\,\mu\,{V}
\,{W}+\nonumber\\
&{\frac {{\beta(t)}\,\mu\, \left( {\alpha_0}+{\gamma_0}
 \right) {{\left (-\frac{\gamma_0^2}{\left (\alpha_0 +\gamma_0\right )^2}W\right )}}^{2}}{{\gamma_0}}}+\mu\sigma\left ( \frac{(\alpha_0 +\gamma_0)}{\gamma_0}U+W+I_0\right )\phi.
\end{flalign}
\end{subequations}
\end{footnotesize}

\newpage

\end{document}